\documentclass[12pt]{iopart} 
\usepackage{iopams}   
\usepackage{graphicx,cite,color,epsf} 
 
\begin{document} 

\title[Universal features of complex $n$-block copolymers]{Universal features of complex $n$-block copolymers} 

\author{K. Haydukivska} 
\address{Institute for Condensed Matter Physics of the National Academy of Sciences of Ukraine, 
79011 Lviv, Ukraine} 
\author{V. Blavatska} 
\address{Institute for Condensed Matter Physics of the National Academy of Sciences of Ukraine, 
79011 Lviv, Ukraine} 
\ead{viktoria@icmp.lviv.ua}

\begin{abstract} 
We study the conformational properties of complex polymer macromolecules, consisting in general of $n$ subsequently
connected chains (blocks)
of different lengths and distinct chemical structure. Depending on the solvent conditions,
the inter- or intrachain interactions
of some blocks may vanish, causing rich conformational behavior.
Our main attention is focused on the universal conformational properties of
such molecules.   Applying the direct polymer renormalization group approach, we derive the analytical expressions for
the scaling exponent $\gamma(n)$, governing the number of possible conformations of $n$-block copolymer, and analyze
the effective linear size measures of individual blocks. In particular,  the degree of extension of the block sizes as functions of $n$
and position of blocks in sequence is quantitatively estimated. The numerical simulations of the simplest $n=2$-block copolymer chain on cubic lattice are performed as
well to give an illustration of the conformational behavior of such molecules.

\end{abstract} 


\submitto{\JPA} 

\section{Introduction}

Block copolymers are the specific class of polymer macromolecules,
containing the subsequent blocks of chemically distinct monomers \cite{Hadjichristidis03}.
The ubiquitous feature of block-copolymer melts is their ability to  microphase separation and
self-assembling into lamellae, micelles and more complex structures \cite{Matsen96,Bates99,Mai12},
which provides potential or practical applications in many fields.
In particular, block copolymers are used for developing dense and nanoporous membranes for gas separation
and ultrafiltration \cite{Jackson10,Dami17,OssRonen12},
in targeted drug delivery  \cite{Meng09}, lithography \cite{Ruiz06}, development of novel
plastic materials \cite{Bates01}, nanotechnology \cite{Schutz05} etc.

The most basic system is the diblock copolymer (so called AB-copolymer):  the molecule 
composed of just two chemically distinct linear blocks denoted A and B. Such systems have been investigated
most extensively. Extension to the multiblock copolymers with $n$ blocks of $k$ chemically
distinct types as well as to more complex branched architectures
 can produce much richer self-assembling behavior.
The modern  synthesis strategies and
techniques, e.g. controlled polymerization technique, allow to produce copolymers with
controlled molecular weights and architectures \cite{Uhrig02}.
For recent reviews on complex copolymers see e.g. \cite{Feng17,Bates12}.

\begin{figure}[b!]
\begin{center}
\includegraphics[width=60mm]{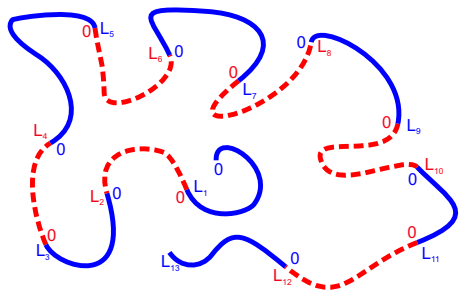}
\end{center}
\caption{Schematic presentation of complex polymer macromolecule, consisting of chemically distinct subsequent blocks. } \label{fig:1}
\end{figure}


Note that whereas mainly the melts of copolymer molecules attract 
interest from technological point of view, in our study we  are concentrated on the 
statistical conformational properties of individual molecules in a very 
dilute regime. This allows, for example, to determine the length scale of 
the block copolymer and relate it to the length scale of aggregates in more concentrated 
solutions or in the membranes \cite{Dami17}. { The conformational properties of diblock AB 
copolymers in dilute regime served as a subject of intensive analytical \cite{Joanny84,Douglas87,McMullen89} 
and numerical \cite{Tanaka76,Tanaka77,Tanaka79,Molina94,Olaj98a} studies. More complex triblock and multiblock copolymers were analyzed in Refs. \cite{Sdranis91,Olaj98b,Olaj98c,Theodorakis12}. }


It is established, that in statistical description of long flexible polymer chains
one finds a number of characteristics which are universal, i.e. not dependent on any details of chemical microstructure.
In particular, the averaged end-to-end distance of a chain of total length (molecular weight) $L$ scales according to \cite{deGennes,desCloiseaux}
\begin{equation}
\langle R^2\rangle \sim L^{2\nu}\label{R},
\end{equation}
where $\langle (\ldots) \rangle$ denotes averaging over an ensemble of all possible conformations which attains the molecule in space,
and $\nu$ is universal scaling exponent. For  the case,
when polymer is in solution at the $\theta$-temperature regime when attractive and repulsive interactions
between monomers are balanced, {the scaling (\ref{R}) holds with $\nu_{\Theta}=4/7$  in $d=2$ dimensions \cite{Duplantier87}. At $d=3$, which is an upper critical
dimension of $\Theta$-transition, one observes the mean-field behavior  $\nu_{\Theta}(d=3)=1/2$ except for the logarithmic corrections 
 \cite{Duplantier86,Duplantier86a}. In the $\Theta$-regime, the polymer chains are 
fairly adequately described by a model of  a freely
jointed chain, similar to that of a diffusing particle executing
a random flight. Here, each the skeletal bond of the chain
resembles a step in the random walk (RW) trajectory, 
whose averaged end-to-end distance scales with
  $\nu_{RW}=1/2$. This corresponds to the regime of ideal Gaussian polymer in continuous chain description of macromolecule \cite{desCloiseaux}.
     The model of random walk is widely exploited in studying the conformational properties of  polymers in $\Theta$-regime in $d=3$  \cite{deGennes,Properties,Polymerization}.}
For a chain in a regime of good solvent (with repulsive excluded volume interactions between monomers playing the role),
 the polymer chain behaves as a trajectory of self-avoiding random walk (SAW), which is not allowed to cross itself.  
  In particular, a good approximation is given by empirical Flory formula:
   $\nu_{SAW}(d)=3/(d+2)$   \cite{deGennes},
  which is a nice agreement with results of more refined studies:
   $\nu_{SAW}(d=2)=3/4$ \cite{Nienhuis82}, $\nu_{SAW}(d=3)= 0.58759700(40)$ \cite{Clisby16}.

The number of possible conformations (partition function) of polymer chain is given by:
\begin{equation}
Z(L)\sim\mu^{L}L^{\gamma-1}, \label{stat}
\end{equation}
here $\mu$ is a non-universal { connectivity constant}  and $\gamma$ is another universal scaling exponent.
{
At the $\theta$-temperature regime, one has $\gamma_{RW}=1$, whereas for a chain in a regime of good solvent:
$\gamma_{SAW}(d=2)=43/32$ \cite{Nienhuis84},  $\gamma_{SAW}(d=3)=1.15695300(95)$ \cite{Clisby17}.
}

Linking together two linear polymer chains of distinct chemical structure A and B, we receive the
simplest diblock copolymer. 
Depending on the temperature, a situation may occur
when interaction within one block
vanishes. Such a block can effectively be described as RW (Gaussian chain), whereas the other behaves as SAW at the same temperature. On the other hand,
also the intrachain interactions
between monomers of different blocks can be taken into account. For example, one can have only the
 mutual excluded-volume interactions between
blocks, whereas each block
can intersect itself. Such a situation is closely related to the case of so-called ternary solutions \cite{schafer1,schafer2}, containing chains of two different chemical structures at various temperatures, which can be at or above the $\theta$-temperatures for
 both or only one type of polymer chains. {  It is established \cite{Joanny84,Douglas87,Tanaka76} that there are actually three  characteristic 
length scales in such systems. One should distinguish between the end-to-end distances of two blocks A and B, which scale
according to (\ref{R}) with exponents $\nu_{A}$ and $\nu_{B}$ correspondingly, where $\nu_{A}$ and $\nu_{B}$ are given by either
$\nu_{SAW}$ or $\nu_{RW}$. The total mean-squared  end-to-end distance of diblock copolymer chain as given by 
 $\langle R^2 \rangle=\langle (\vec{R}_A+ \vec{R}_{B})^2\rangle $, displays non-trivial behaviour when mutual interactions between blocks $A$ and $B$ are present. 
  }  
 
Partition sum of the copolymer chain consisting of blocks with lengths $L_1$ and $L_2$ scales as:
\begin{equation}
Z(L_1,L_2)\sim \mu_1^{L_1}\mu_2^{L_2}(L_1+L_2)^{\gamma-1},\label{stat2}
\end{equation}
where $\mu_1$, $\mu_2$ are { connectivity constants} of corresponding blocks.
 {The universality class (set of scaling exponents) $\gamma$ of diblock copolymers is actually given by that of polymer chains in ternary solutions
 \cite{schafer1,schafer2,Holovatch97}.
 The estimates of scaling exponents $\gamma$ governing the scaling (\ref{stat2}) can be derived e.g. on the basis of results 
for more general branched copolymer structures obtained within the field-theory renormalization group technique in Ref. \cite{Holovatch97}. On the other hand, the scaling exponent $\gamma$ governing the scaling of a set of mutually avoiding Gaussian chains have been estimated  in a number of works
dedicated to probability of intersection of Brownian paths \cite{Duplantier88,Duplantier88a,Lawler82,Lawler90,Li90}.}

Generalizing expression (\ref{stat2}) for the case of $n$-block copolymer with segments of different lengths $L_i$ we have:
 \begin{equation}
Z(L_1,\ldots,L_n)\sim \prod_{i=1}^n\mu_i^{L_i}\left(\sum_{i=1}^nL_i\right)^{\gamma(n)-1}. \label{stat3}
\end{equation}
Partition function depends on $\mu_i$ of 
 each block, which  can differ 
from  that of individual polymer chains
 due to crowdedness effect, caused by presence of other 
blocks. 
Note, that in the limit of  large $n$, one can in principle 
replace each $\mu_i$ with the averaged value 
$\mu_{av}=1/n \sum_i\mu_i$ and restore  Eq. (1) with 
$L=\sum_iL_i$ and $\mu=\mu_{av}$.

 The aim of the present work is to analyze  the conformational properties of complex copolymers
 consisting of $n$ blocks of different lengths $L_i$, $i=1,\ldots,n$ (Fig. \ref{fig:1}) in {$d=3$, applying analytical approach of 
 direct polymer renormalization}.  The layout of the paper is as follows.  In Section \ref{model} we present
 the continuous chain model of $n$-block copolymer macromolecule. In Section \ref{diblock} we derive the analytical description of properties of
 diblock copolymer (with $n=2$), and generalize it to the $n$-block case
in Section (\ref{nblock}). In Section \ref{num}, we present some results of  numerical simulations of diblock copolymer chain
 on simple cubic lattice using the algorithm of growing chain.  We conclude with discussions in Section \ref{Discus}.

\section{Analytical studies of $n$-block copolymer}

\subsection{Continuous chain model}\label{model}

Within the continuous chain model, each  block of the copolymer chain is considered as a trajectory parameterized by radius vector $\vec{r}_i(s)$ with $s$ changing from $0$ to $L_i$, $i=1,\ldots,n$.  The $n$-block copolymer can be thus presented as a set of $n$ trajectories,
consequently connected by their end points.
The partition sum of the system is thus given by
\begin{eqnarray}
Z(L_1,\ldots,L_n)= \int {\cal D}\, \vec{r}  \prod_{i=1}^{n-1}\delta(\vec{r}_i(L_i)-\vec{r}_{i+1}(0))\, {\rm e}^{-H}, \label{Z}
\end{eqnarray}
here $\int {\cal D}\, \vec{r}$ denotes functional integration over $n$ trajectories and
 $H$ is an effective Hamiltonian of the system:
\begin{eqnarray}
&&H=\frac{1}{2}\sum_{i=1}^n\int_0^{L_i}{\rm d} s \left(\frac{d\vec{r}_i(s)}{ds}\right)^2+\sum_{i=1}^n u_i\int_0^{L_i}\!\!{\rm d}s_1\int_0^{L_i}\!\!{\rm d}s_2 \,\delta(\vec{r}_i(s_2)-\vec{r}_i(s_1))\nonumber\\
&&+\sum_{i<j=1}^n w_{i,j}\int_0^{L_i}\!\!{\rm d}s_1\int_0^{L_j}\!\!{\rm d}s_2 \,\delta(\vec{r}_i(s_2)-\vec{r}_j(s_1)).
\label{H}
\end{eqnarray}
Here, $u_i$ are the coupling constants of excluded volume interactions between  monomers of the same block
 and $w_{i,j}$ correspond to interactions between different blocks. Note that in the absence of interactions ($u_i=w_{i,j}=0$) 
we just restore the case of idealized Gaussian chain.  

  Evaluating the perturbation theory expansion in coupling constants $u_i$ and $w_{i,j}$
and limiting ourself only  to the first order approximation, we obtain an expression
\begin{eqnarray}
&&Z(L_1,\ldots,L_n)=1+\sum_{i=1}^n \frac{u_i}{(2\pi)^{d/2}} \int_0^{L_i}\!\!{\rm d}s_1\int_0^{L_i}\!\!\!{\rm d}s_2(s_2-s_1)^{-\frac{d}{2}}\\
&&+\sum_{i<j=1}^n\frac{w_{i,j}} {(2\pi)^{d/2}}\int_0^{L_i}\!\!\!{\rm d}s_1\int_0^{L_j}\!\!{\rm d}s_2 (s_1+s_2+l_{i,j})^{-\frac{d}{2}},\nonumber
\end{eqnarray}
where $l_{i,j}=\sum_{t=i+1}^{j-1}L_{t}$. Performing the integration, we have:
\begin{eqnarray}
&&Z(L_1,\ldots,L_n)=1-\sum_{i=1}^n \frac{u_i}{(2\pi)^{d/2}} \frac{L_i^{2-d/2}}{(1-d/2)(2-d/2)}\nonumber\\
&&-\sum_{i<j=1}^n \frac{w_{i,j}}{(2\pi)^{d/2}}\left(\frac{(l_{i,j})^{2-d/2}-(L_j+l_{i,j})^{2-d/2}}{(1-d/2)(2-d/2)}\right.\nonumber\\
&&\left.+\frac{(L_i+L_j+l_{i,j})^{2-d/2}-(L_i+l_{i,j})^{2-d/2}}{(1-d/2)(2-d/2)}\right). \label{Zexp}
\end{eqnarray}
{ Generalized form of the scaling behavior of partition sum of $n$-block copolymer is given by Eq. (\ref{stat3}).}

Let us note, that performing the dimensional analysis of couplings, we find that $[u_i]=[L_i]^{(4-d)/2}$, $[w_{i,j}]=[(L_i+L_j)]^{(4-d)/2}$. The ``upper critical'' value
of space dimension $d_c=4$, at which the coupling becomes dimensionless, plays the role in the following application of renormalization group scheme.
{ It is important to stress, that though all the analytical expressions in the rest of this Section are derived  
in a form of series in deviation from the upper-critical dimension $\epsilon = 4-d$, which is standard in renormalization group approach, 
we are interested only in the case $d=3$ (evaluating the final results for quantities of interest at $\epsilon=1$). }

Within our approach, the value of scaling exponent $\gamma$ can be found according to \cite{desCloiseaux}:
\begin{eqnarray}
&&\gamma(n)= \frac{\epsilon}{2}\left(\sum_{i=1}^n\frac{\partial \ln Z(L_1,\ldots,L_n)}{\partial\ln \widetilde{u}_i}\right.\left.+\sum_{i<j=1}^n\frac{\partial \ln Z(L_1,\ldots,L_n)}{\partial \ln\widetilde{w}_{i,j}}\right)\label{gamma}
\end{eqnarray}
where $\widetilde{u}_i$ and $\widetilde{w}_{i,j}$  are dimensionless coupling constants:
\begin{eqnarray}
&&\widetilde{u}_i= \frac{u_i}{(2\pi)^{d/2}}L_i^{2-d/2},\nonumber\\
&&\widetilde{w}_{i,j} = \frac{w_{i,j}}{(2\pi)^{d/2}} (L_i+L_j)^{2-d/2}.\label{dimu}
\end{eqnarray}

\subsection{Diblock copolymer chain}\label{diblock}

{\bf Partition sum.} Let us start with the most simple case, when the system consists of only two connected block chains  of lengths $L_1$ and $L_2$. For the partition function, we have:
 \begin{eqnarray}
&&Z(L_1,L_2)=1- \frac{u_1}{(2\pi)^{d/2}} \frac{L_1^{2-d/2}}{(1-d/2)(2-d/2)}\nonumber\\
&&-\frac{u_2}{(2\pi)^{d/2}}\frac{L_2^{2-d/2}}{(1-d/2)(2-d/2)}\\
&&-\frac{w_{1,2}}{(2\pi)^{d/2}} \frac{(L_1+L_2)^{2-d/2}-(L_2)^{2-d/2}- (L_1)^{2-d/2}}{(1-d/2)(2-d/2)}.\nonumber
\end{eqnarray}

Presenting the partition function in a form of series in  $\epsilon$
and passing to the dimensionless coupling constants according to (\ref{dimu}),
we receive:
\begin{eqnarray}
&&Z(L_1,L_2)=1+\left(\widetilde{u}_1+\widetilde{u}_2-\widetilde{w}_{1,2}\right)\frac{2}{\epsilon}\nonumber\\
&&+\widetilde{u}_1+\widetilde{u}_2-\widetilde{w}_{1,2}\left(1+\ln\left(\frac{L_1L_2}{(L_1+L_2)^2}\right)\right).
\end{eqnarray}

{\bf End-to end distance.}
 There are two characteristic length scales in diblock copolymer, given by
end-to-end distances:
\begin{eqnarray}
&&\langle R^2_{1}\rangle = \langle (\vec{r}_1(L_1)-\vec{r}_1(0))^2\rangle\\
&&\langle R^2_{2}\rangle = \langle (\vec{r}_2(L_2)-\vec{r}_2(0))^2\rangle,
\end{eqnarray}
where indexes $1$ and $2$ indicate blocks of copolymer chain.
Performing the perturbation theory expansion in coupling constants, we receive:
\begin{eqnarray}
&&\langle R^2_{1}\rangle =  dL_1\left(1+\frac{u_1}{(2\pi)^{d/2}} \frac{L_1^{2-d/2}}{(3-d/2)(2-d/2)}\right.\nonumber\\
&&\left. + \frac{4w_{1,2}}{(2\pi)^{d/2}}\left(\frac{(L_1+L_2)^{-d/2}}{L_1}\frac{dL_1^3+d^2L_1^2L_2-6dL_1^3}{d(d-6)(d-4)(d-2)}\right.\right.\nonumber\\
&&+\frac{(L_1+L_2)^{-d/2}}{L_1}\frac{4dL_1L_2^2-2dL_1^2L_2+8(L_1^3+L_2^3)}{d(d-6)(d-4)(d-2)}\nonumber\\
&&\left.\left.+\frac{1}{L_1}\frac{d(6-d)L_1^{3-d/2}-8\left(L_1^{3-d/2}+L_2^{3-d/2}\right)}{d(d-6)(d-4)(d-2)}\right)\right),\nonumber\\
&&\langle R^2_{2}\rangle = dL_2\left(1+\frac{u_2}{(2\pi)^{d/2}} \frac{L_2^{2-d/2}}{(3-d/2)(2-d/2)}\right.\nonumber\\
&&\left.+ \frac{4w_{1,2}}{(2\pi)^{d/2}}\left(\frac{(L_1+L_2)^{-d/2}}{L_2}\frac{dL_2^3+d^2L_2^2L_1-6dL_2^3}{d(d-6)(d-4)(d-2)}\right.\right.\nonumber\\
&&+\frac{(L_1+L_2)^{-d/2}}{L_2}\frac{4dL_2L_1^2-2dL_2^2L_1+8(L_1^3+L_2^3)}{d(d-6)(d-4)(d-2)}\nonumber\\
&&\left.\left.+\frac{1}{L_2}\frac{d(6-d)L_2^{3-d/2}-8\left(L_1^{3-d/2}+L_2^{3-d/2}\right)}{d(d-6)(d-4)(d-2)}\right)\right).
\end{eqnarray}
As for the case of partition function, we pass  here to the dimensionless coupling constants and
perform the series expansions in  $\epsilon$:
\begin{eqnarray}
&&\langle R^2_{1}\rangle = dL_1\left(1+\widetilde{u}_1\frac{2}{\epsilon}-\widetilde{u}_1-\frac{\widetilde{w}_{1,2}}{2(L_1+L_2)}\times\right.\nonumber\\
&&\times\left.\frac{L_2}{L_1}\left(2(L_1+L_2)\ln\left(\frac{L_2}{L_1+L2}\right)+L_1\right)\right), \label{R1}\\
&&\langle R^2_{2}\rangle = dL_2\left(1+\widetilde{u}_2\frac{2}{\epsilon}-\widetilde{u}_2-\frac{\widetilde{w}_{1,2}}{2(L_1+L_2)}\times\right.\nonumber\\
&&\times\left.\frac{L_1}{L_2}\left(2(L_1+L_2)\ln\left(\frac{L_1}{L_1+L2}\right)+L_2\right)\right).\label{R2}
\end{eqnarray}

 Note, that our expressions for $\langle R^2_{1}\rangle$, $\langle R^2_{2}\rangle$ restore corresponding results obtained previously in \cite{Joanny84,Douglas87}.

{\bf Fixed points, scaling exponents and size ratios}

Applying the direct polymer renormalization scheme, briefly given  in Appendix, we obtain the values of FPs governing
the behaviour of different types of block copolymers as common zeros of functions (\ref{b1}), (\ref{b2}). We found 6 sets of FPs, 
which are usually distinguished 
in studies concerning the diblock copolymers \cite{schafer1,schafer2,Holovatch97}:

\begin{eqnarray}
&&1)\, \widetilde{u}_{1,R}= 0,\quad\! \widetilde{u}_{2,R} =0, \quad\! \widetilde{w}_{1,2,R} =0,\label{FP1}\\
&&2)\, \widetilde{u}_{1,R}= 0,\quad\! \widetilde{u}_{2,R} =0,\quad\! \widetilde{w}_{1,2,R} =\frac{\epsilon L_1L_2}{(L_1+L_2)^2},\\
&&3)\, \widetilde{u}_{1,R}= 0,\quad\! \widetilde{u}_{2,R} =\frac{\epsilon}{8}, \quad\! \widetilde{w}_{1,2,R} =0,\\
&&4)\, \widetilde{u}_{1,R}= 0,\quad\!\! \widetilde{u}_{2,R} =\frac{\epsilon}{8}, \quad\!\!\! \widetilde{w}_{1,2,R} =\frac{3\epsilon L_1L_2}{4(L_1+L_2)^2},\\
&&5)\, \widetilde{u}_{1,R}= \frac{\epsilon}{8},\quad\! \widetilde{u}_{2,R} = \frac{\epsilon}{8}, \quad\! \widetilde{w}_{1,2,R} =0,\\
&&6)\, \widetilde{u}_{1,R}= \frac{\epsilon}{8},\quad\!\!\! \widetilde{u}_{2,R} =\frac{\epsilon}{8}, \quad\!\!\! \widetilde{w}_{1,2,R} =\frac{\epsilon L_1L_2}{2(L_1+L_2)^2}.\label{FP6}
\end{eqnarray}

The case (1) corresponds to most trivial situation of two idealized Gaussian chains, whereas case (2) includes mutual avoidance between them.
Similarly, (3) and (4) describe the situation, when one block is Gaussian chain and the other block feels the excluded volume effect, with and without mutual interactions, correspondingly. Finally, (4) and (5) describe two blocks with an excluded volume effect, again with and without mutual interactions.   
Note, that cases (3) and (4)  are ``degenerate'' in the sense, that we should also take into account the sets with symmetrical change of values
$\widetilde{u}_{1,R}$  and $\widetilde{u}_{2,R}$.

 \begin{table}[h!]
\begin{center}
\caption{Scaling exponents governing the behavior of diblock copolymer evaluated at different fixed points.}
\label{tab}
  \begin{tabular}{ c  c  c  c }
   \hline
$  $ & $\nu_1$ &  $\nu_2$ & $\gamma$  \\ 
\hline
$(1)$ & $1/2$ & $1/2$ & $1$ \\ 
$(2)$ & $1/2$ & $1/2$ & $1-\frac{\epsilon L_1L_2}{(L_1+L_2)^2}$ \\ 
$(3)$ & $1/2$ & $	1/2+\epsilon/16$ & $1+\frac{\epsilon}{8}$ \\ 
$(4)$ & $	1/2$ & $	1/2+\epsilon/16$ & $1+\frac{\epsilon}{8}-\frac{3\epsilon L_1L_2}{4(L_1+L_2)^2}$\\ 
$(5)$ & $	1/2+\epsilon/16$ & $	1/2+\epsilon/16$ & $1+\frac{\epsilon}{4}$\\ 
$(6)$ & $1/2+\epsilon/16$ & $	1/2+\epsilon/16$ & $1+\frac{\epsilon}{4}-\frac{\epsilon L_1L_2}{2(L_1+L_2)^2}$\\ 
\hline
    \end{tabular}
\end{center}
\end{table}

The expression for  $\gamma$ as given by Eq. (\ref{gamma}) in the case of diblock copolymer reads:
\begin{eqnarray}
&&\gamma=1+\left(\widetilde{u}_1+\widetilde{u}_2-\widetilde{w}_{1,2}\right)+\epsilon\left(\widetilde{u}_1+\widetilde{u}_2-\widetilde{w}_{1,2}\right.\nonumber\\
&&\left.-\epsilon\widetilde{w}_{1,2}\ln\left(\frac{L_1L_2}{(L_1+L_2)^2}\right)\right).\label{gamma2}
\end{eqnarray}

Substituting the values of FPs for the cases (1)-(6) into Eq. (\ref{gamma2})  and keeping  terms only up to the linear order in $\epsilon$, we obtain the corresponding values of
scaling exponents $\gamma$ given in Table (\ref{tab}).

Let us consider situation when $L_1=L_2$. 
{ In this case, we can restore the first-order expansion for $\gamma$ from the set of star copolymer exponents $\eta_{f_1,f_2}$ obtained in Ref. \cite{Holovatch97}.  We can make use of scaling relation, which takes place for diblock copolymer chain: 
$ \gamma^{(i)}=1+\nu^{(i)}(\eta^{(i)}_{1,1}-\eta^{(i)}_{2,0}-\eta^{(i)}_{0,2})$, where $\eta_{2,0}^{(SAW)}=\eta_{0,2}^{(SAW)}=(1-\gamma^{SAW})/\nu^{SAW}=-\epsilon /4$,
$\eta_{2,0}^{(RW)}=\eta_{0,2}^{(RW)}=0$. Taking the values $\eta_{1,1}^{(2)}=-\epsilon/2$, $\eta_{1,1}^{(4)}=-3\epsilon/8$ from
Ref. \cite{Holovatch97}, we restore 
the values of $\gamma^{(2)}$, $\gamma^{(4)}$ obtained by us.}

Aiming to find the quantitative estimates in this case for $d=3$, we put  $\epsilon=1$ in corresponding expressions in Table \ref{tab}.
 Thus, we have:
\begin{eqnarray}
&&\gamma^{(1)}=1,\\
&&\gamma^{(2)}=\frac{3}{4},\\
&&\gamma^{(3)}=\frac{9}{8},\\
&& \gamma^{(4)}=\frac{15}{16},\\
&& \gamma^{(5)}=\frac{5}{4},\\
&&\gamma^{(6)}=\frac{9}{8}.
\end{eqnarray}

{
For comparison, we present the corresponding values obtained on the basis of the 3rd order of $\epsilon$ expansion given in Ref. \cite{Holovatch97} for 
the set of star copolymer exponents $\eta_{f_1,f_2}$.  Again, making use of the scaling relation mentioned above and taking $\eta_{2,0}^{(SAW)}=\eta_{0,2}^{(SAW)}=-0.2671$,
$\eta_{2,0}^{(RW)}=\eta_{0,2}^{(RW)}=0$, $\eta_{1,1}^{(2)}=-0.46$, $\eta_{1,1}^{(4)}=-0.28$ (at $\epsilon=1$) we obtain: $\gamma^{(2)}=0.770$, $\gamma^{(4)}=0.992$.
}

We can also estimate the values of exponents $\nu_i$, governing correspondingly
the scaling behavior of the end-to-end distances $\langle R_1^2\rangle$ and $\langle R_2^2\rangle$ of two blocks. We receive \cite{desCloiseaux}
\begin{eqnarray}
&&\nu_i=\frac{\epsilon}{2}\sum_{i=1}^2\widetilde{u}_i\frac{{\rm d} \ln \langle R_i^2\rangle}{{\rm d}\ln\widetilde{u}_i}+\widetilde{w}_{1,2}
\frac{{\rm d} \ln \langle R_i^2\rangle}{{\rm d}\ln\widetilde{w}_{12}}.\label{nu}
\end{eqnarray}
Substituting the expressions (\ref{R1}) and (\ref{R2}) into this relation, we obtain:
\begin{eqnarray}
&&\nu_1=\frac{1}{2}\left(1+\widetilde{u}_1-\epsilon\widetilde{u}_1+\frac{\epsilon\widetilde{w}_{1,2}}{2(L_1+L_2)}\times\right.\nonumber\\
&&\times\left.\frac{L_2}{L_1}\left(2(L_1+L_2)\ln\left(\frac{L_2}{L_1+L_2}\right)+L_1\right)\right),\label{nu1}\\
&&\nu_2=\frac{1}{2}\left(1+\widetilde{u}_2-\epsilon\widetilde{u}_2+\frac{\epsilon\widetilde{w}_{1,2}}{2(L_1+L_2)}\times\right.\nonumber\\
&&\times\left.\frac{L_1}{L_2}\left(2(L_1+L_2)\ln\left(\frac{L_1}{L_1+L_2}\right)+L_2\right)\right).\label{nu2}
\end{eqnarray}
Substituting the values of FPs for the cases (1)-(6) into Eqs. (\ref{nu1}) and (\ref{nu2}) and keeping again terms only up to the linear order in $\epsilon$, we obtain the corresponding values of
scaling exponents $\nu_1$, $\nu_2$ given in Table (\ref{tab}).
Note that the values of these exponents  do not depend on $L_1$, $L_2$ and, mostly important, they are not modified by presence of mutual avoidance with other block chain. 

However, the presence of interaction with other block modifies the values of $\langle R_1^2\rangle$ and $\langle R_2^2\rangle$ as compared with that of a free chain. This can be easily seen by introducing the ratios:
\begin{eqnarray}
g_i=\frac{\langle R^2 \rangle_{chain}}{\langle R^2_i \rangle},\label{rat}
\end{eqnarray}
where $\langle R^2 \rangle_{chain}$ is the end-to-end distance of a single chain of a length $L$:
\begin{eqnarray}
&&\langle R^2_{chain}\rangle = dL\left(1+\widetilde{u}_1\frac{2}{\epsilon}-\widetilde{u}_1\right).\label{Rchain}
\end{eqnarray}
Thus, we have:
\begin{eqnarray}
&&g_1 = 1 +\frac{\widetilde{w}_{1,2}}{2(L_1+L_2)}\frac{L_2}{L_1}\left(L_1+2(L_1+L_2)\ln\left(\frac{L_2}{L_1+L2}\right)\right), \\
&&g_2 = 1 +\frac{\widetilde{w}_{1,2}}{2(L_1+L_2)}\frac{L_1}{L_2}\left(L_2+2(L_1+L_2)\ln\left(\frac{L_1}{L_1+L2}\right)\right).
\end{eqnarray}
Again, considering the case when $L_1=L_2$ we have:
  \begin{eqnarray}
&&g_i = 1 - \widetilde{w}_{1,2}(\ln(2)-1/4). \label{RRR1}
\end{eqnarray}
Substituting the FP value of $w_{12}$ for the cases (1)-(6), one can easily see that in presence of mutual avoidance between blocks
 the ratio is always smaller than 1 and thus the effective size of a block is more extended in space.

\subsection{$n$-block copolymer chain}\label{nblock}

{\bf End-to-end distances}
The previous description can be easily generalized to a more complicated polymer  consisting of $n$
subsequently connected blocks. So, we come to the problem  with the set of $n$ coupling constants governing the excluded volume interactions
within the same block ($u_i$ with $i=1,\ldots,n$) and $n(n-1)/2$ coupling constants governing the mutual interactions between  any pair
of blocks ($w_{i,j}$ with $i<j=1,\ldots,n$).  Correspondingly, there are $n$ characteristic lengths in a system, given by:
\begin{eqnarray}
&&\langle R^2_{i}\rangle = dL_i\left(1+\widetilde{u}_i\left(\frac{2}{\epsilon}-1\right)+\right.
\sum_{j \neq i =1}^n\widetilde{w}_{i,j} \left[\frac{l_{i,j}}{L_i}\ln\left(\frac{l_{i,j}}{L_i+l_{i,j}}\right)\right.\nonumber\\
&&-\frac{l_{i,j}+L_j}{L_i}\ln\left(\frac{l_{i,j}+L_j}{L_i+l_{i,j}+L_j}\right)\left.-\frac{L_iL_j}{2(L_i+l_{i,j}+L_j)(L_i+l_{i,j})}\right]\\
&&\left.+\sum_{k=1}^{i-1}\sum_{m=i+1}^n\frac{\widetilde{w}_{k,j}L_kL_j(2l_{kj}+L_k+L_j)}{2(L_k+l_{kj}+L_j)(L_k+l_{kj})(l_{kj}+L_j)}\right),\nonumber
\end{eqnarray}
with  $l_{i,j}=\sum_{t=i+1}^{j-1}L_{t}$, and it is important to note that  $l_{km}$ depends on $L_i$ by containing it in the sum.

Again, we can consider the size ratio (\ref{rat}) to compare the effective size of $i$th block in $n$-block copolymer chain with that of
  a single chain of the same length $L_i$.
  The result reads:
\begin{eqnarray}
&&g_i = 1-\sum_{j \neq i =1}^n\widetilde{w}_{i,j} \left[\frac{l_{i,j}}{L_i}\ln\left(\frac{l_{i,j}}{L_i+l_{i,j}}\right)\right.
-\frac{l_{i,j}+L_j}{L_i}\ln\left(\frac{l_{i,j}+L_j}{L_i+l_{i,j}+L_j}\right)\nonumber\\
&&\left.-\frac{L_iL_j}{2(L_i+l_{i,j}+L_j)(L_i+l_{i,j})}\right)\nonumber\\
&&-\sum_{k=1}^{i-1}\sum_{m=i+1}^n \frac{\widetilde{w}_{km}L_kL_m(2l_{km}+L_k+L_m)}{2(L_k+l_{km}+L_m)(L_k+l_{km})(l_{km}+L_m)}.\label{gblock}
\end{eqnarray}

\begin{figure}[t!]
\begin{center}
\includegraphics[width=80mm]{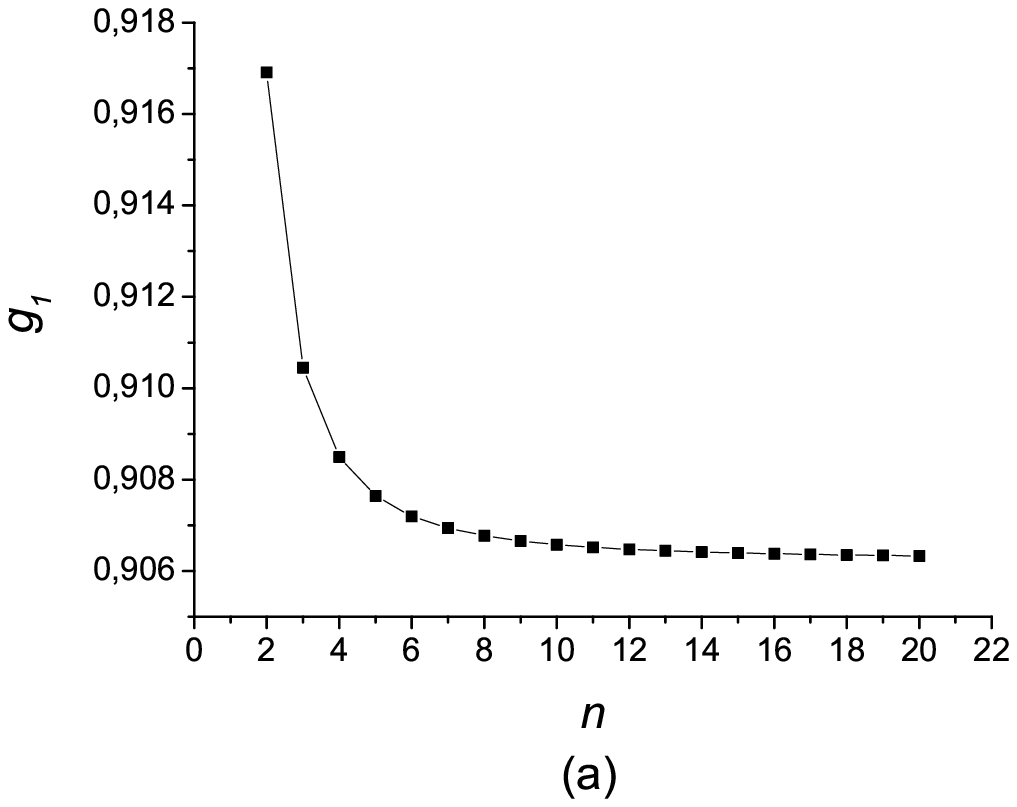}
\includegraphics[width=80mm]{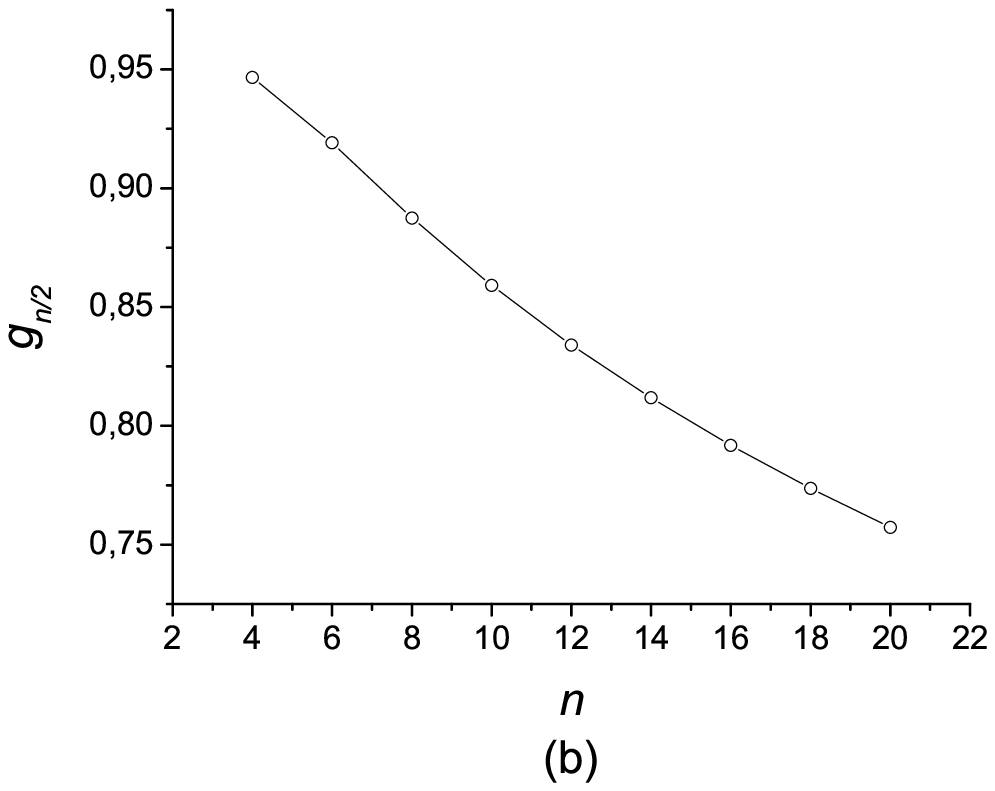}
\end{center}
\caption{The size ratios (\ref{gi}) for the end block (a) and central block (b) as functions of the number of blocks $n$.} \label{Rblock}
\end{figure}

{\bf Fixed points, scaling exponents and size ratios}
Generalizing the direct polymer renormalization group scheme to the system of $n$ blocks,
we come to the system of $n+n(n-1)/2$ flow equations:
\begin{eqnarray}
&&\beta_{u_{i,R}} =\epsilon \widetilde{u}_{i,R}-8\widetilde{u}^2_{i,R}=0, \label{bn1}\\
&&\beta_{w_{i,j,R}} =\epsilon \widetilde{w}_{i,j,R}-\frac{(L_i+L_j)^2}{L_iL_j}\widetilde{w}^2_{i,j,R}
 -2w_{i,j,R}(u_{i,R}+u_{j,R})=0. \label{bn2}
\end{eqnarray}
Analyzing the set of FPs, which are the common zeros of functions (\ref{bn1}), (\ref{bn2}),
we find that $\widetilde{u_{i}}$ can take the two possible values:
\begin{eqnarray}
&&\widetilde{u}_{i,R} = 0,\,\,\,\widetilde{u}_{i,R} = \frac{\epsilon}{8},
\end{eqnarray}
so that the size scaling exponents $\nu_i$, governing the scaling of the end-to-end distances $\langle R^2_i \rangle$
again take on the values of either $1/2$ or $1/2+\epsilon/16$, correspondingly.

For the $\widetilde{w}_{i,j}$ we have:
\begin{eqnarray}
&&\widetilde{w}_{i,j,R} = 0, \, \forall \,\widetilde{u}_i, \, \widetilde{u}_j, \label{1}\\
&&\widetilde{w}_{i,j,R} = \frac{\epsilon L_iL_j}{(L_i+L_j)^2},\,\widetilde{u}_i=\widetilde{u}_j=0,\label{2} \\
&&\widetilde{w}_{i,j,R} = \frac{3\epsilon L_iL_j}{4(L_i+L_j)^2},\, \widetilde{u}_i\neq \widetilde{u}_j, \label{3}\\
&&\widetilde{w}_{i,j,R} = \frac{\epsilon L_iL_j}{2(L_i+L_j)^2},\, \widetilde{u}_i=\widetilde{u}_j= \frac{\epsilon}{8}.\label{4}
\end{eqnarray}

Applying formula (\ref{gamma}) to the expression for the partition function of $n$-block copolymer (\ref{Zexp}) and expanding the result into series over $\epsilon$ with keeping only the terms that are linear in coupling constants we receive:
\begin{eqnarray}
&&\gamma(n)=1+\sum_{i=1}^n \widetilde{u}_i - \sum_{i=1}^{n-1}\widetilde{w}_{i,i+1}.
\end{eqnarray}
It is interesting to note that in one-loop approximation $\gamma(n)$ depends only on interactions between neighboring segments,
 whereas the partition function depends on all $\widetilde{w}_{i,j}$.

For an illustration, let us consider the
alternating sequence of the type ABABA..., where A and B are correspondingly blocks of the SAW type
(with $u_{2i+1}=\epsilon/8$, $i=0,\ldots,n/2-1$)
and RW type (with $u_{2i}=0$, $i=1,\ldots,n/2$), all of equal length.
In the case, when there are no mutual interactions between blocks ($w_{i,j}=0$),
we have:
\begin{equation}
\gamma(n)=1+\frac{n\epsilon}{16}.
\end{equation}
When mutual interactions are present between all pairs of blocks ($w_{i,i+1}=3\epsilon/16$) we obtain:
\begin{equation}
\gamma(n)=1+\frac{n\epsilon}{16}-(n-1)\frac{3\epsilon}{16}.
\end{equation}
{ Let us remind, that aiming to describe the 3-dimensional system, we should put $\epsilon=1$ in this expression.}
 
For the size ratio (\ref{gblock}) in this situation we have:
\begin{eqnarray}
&&g_i=1-\frac{3\epsilon}{16}\left[\sum_{j \neq i =1}^n \left( (|j-i|-1)\ln\left ( \frac{|j-i|-1}{|j-i|}\right) \nonumber \right.\right. \\
&&-\left.|j-i|\ln\left(\frac{|j-i|}{1+|j-i|} \right) -\frac{1}{2(1+|j-i|)|j-i|} \right)\nonumber\\
&& \left.-\sum_{k=1}^{i-1}\sum_{m=i+1}^n \frac{1}{(m-k+1)(m-k)}\right]. \label{gi}
\end{eqnarray}

It is important to note, that the effective linear size measures of blocks depend not only on the overall number of blocks $n$, but also on the
position $i$ of the block along the sequence. On  Fig. \ref{Rblock}a, we present the results for the end block (with $i=1$, which coincide with the case $i=n$).
At $n=2$, we restore the size ratio (\ref{RRR1}) of the diblock copolymer for the case (6). We note, that $g_1$ slightly decreases with growing of $n$, so that the block becomes more extended in space as comparing with the size of single polymer chain of the same length, due to presence of another blocks.
This effect becomes much more pronounced, when we analyze the size ratio for the central block (with $i=n/2$), presented on Fig.  \ref{Rblock}b.
Due to a large amount of  contacts with neighboring blocks, the effective size of central block is more extended in space as compared with that of end blocks, and continuously grows  with increasing the total number of blocks $n$.

\section{Numerical simulations of diblock copolymer chain}\label{num}

{To provide some illustrations of our analytical results of the previous Section,  we present here some results 
of the lattice simulations of  diblock copolymer chain on a cubic lattice.
}
We assume, that each block consists of $N$ monomers (conjugated to the continuous length $L$ in Eqs. (\ref{R}), (\ref{stat})).
The following 6 possibilities are considered, directly corresponding to the
cases (\ref{FP1})-(\ref{FP6}) introduced in previous Section:
 
1) Both blocks are RW trajectories without mutual avoiding, so we restore the RW trajectory of total length $2N$

2)  Both are RW trajectories, but avoiding each other: each trajectory can cross itself, but cannot cross the other trajectory

3) One block is SAW, the other is RW, without mutual avoiding

4) One block is SAW, the other is RW, avoiding each other

5) Both are SAW trajectories without mutual avoiding

6) Both are SAW trajectories avoiding each other, so we restore the SAW trajectory of total length $2N$

In our simulations we apply the Rosenbluth-Rosenbluth growing chain algorithm \cite{Rosenbluth55}. Two chains are growing simultaneously
from the same starting point, so that the $n$th monomer of each chain is placed at a randomly chosen neighboring site of the  $(n-1)$th  monomer ($n\leq N$).
Note that some amount of nearest neighbor sites can be forbidden for placing a new monomer, if they are already visited
(depending on conditions (1)-(6)).
The weight $W_n= \prod_{l=1}^n m_1^lm_2^l$ is given to each sample configuration at the $n$th step,
where $m_1^l$ and $m_2^l$ are the numbers of free lattice sites to place the $l$th monomer of the first and second chains, correspondingly.
 The growth is stopped, when the total length $N$ of each the block chain is reached.
 Then the next cofiguration is started to grow from the same starting point.
The configurational averaging of any observable is thus given by:
\begin{eqnarray}
&&\langle (\ldots) \rangle=\frac{\sum_{t=1}^{{\rm conf}}W_N^{{t}}(\ldots)}{\sum_{t=1}^{{\rm conf}}W_N^{t}},
\end{eqnarray}
where ${\rm conf}$ is the number of different configurations constructed.
The partition sum $Z_N$ is thus estimated as an averaged weight:
\begin{equation}
 Z_N  =\frac{1}{{\rm conf}}\sum_{t=1}^{{\rm conf}}W_N^{{t}}.
\label{number}
\end{equation}
We constructed chains with $N$ up to $150$ and performed 100 tours with building $10^6$ configurations in each of the tour.

\begin{figure}[t!]
\begin{center}
\includegraphics[width=80mm]{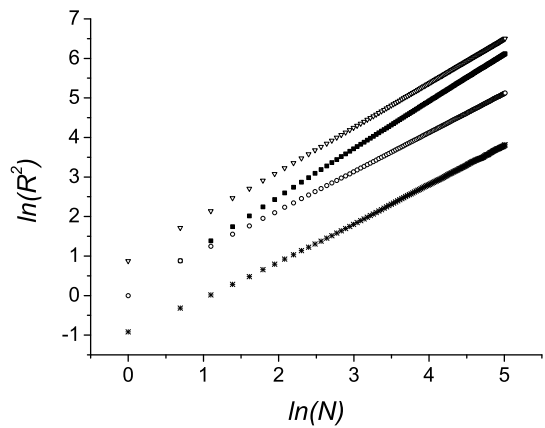}
\end{center}
\caption{The averaged end-to-end distances $\langle R^2 \rangle^{(4)}$ (open triangles), $\langle R^2 \rangle_{SAW}^{(4)}$ (black squares), $\langle R^2 \rangle_{RW}^{(4)} $
 (open circles), and $\langle \vec{R}_{SAW}^{(4)} \vec{R}_{RW}^{(4)}\rangle$ (stars)  as functions of $N$ in double logarithmic scale.} \label{RR}
\end{figure}

We start with  an effective size measure of the two-block copolymer chain.
In the problem under consideration, we have two characteristic length scales of two blocks A and B:  ${\langle R^2 \rangle_{A}^{(i)}}$ and ${\langle R^2 \rangle_{B}^{(i)}}$, with $i=1,\ldots,6$.
Both of them scale according to Eq. (\ref{R}) with exponents $\nu_{A}$ and $\nu_{B}$ correspondingly, which are given either by $\nu_{SAW}$ or $\nu_{RW}$, 
and the scaling is not modified by
 an avoidance with other block. Note however, than in experiments one mainly investigates the total
 effective length of a block copolymer chain
 \begin{equation}
 {\langle R^2 \rangle^{(i)}}={\langle R^2 \rangle_{A}^{(i)}}+{\langle R^2 \rangle_{B}^{(i)}}+2\langle \vec{R}_{A}^{(i)} \vec{R}_{B}^{(i)} \rangle. \label{corr}
 \end{equation}
 Note, that when the mutual interaction between two blocks is absent (cases (1), (3), (5)), the last term in (\ref{corr}) vanishes, since 
 there is no correlations between vectors $\vec{R}_A^{(i)}$ and $\vec{R}_B^{(i)}$ in this case.
 
 In Fig. \ref{RR}, we present  simulation results for the characteristic lengths for the case (4):
${\langle R^2 \rangle^{(4)}}$, ${\langle R^2 \rangle_{SAW}^{(4)}}$ and ${\langle R^2 \rangle_{RW}^{(4)}}$, and also for the corresponding correlation term
$\langle \vec{R}_{SAW}^{(4)}\vec{R}_{RW} ^{(4)} \rangle$.

{
To estimate the critical exponents, we use the linear least-square fits in the form 
\begin{equation}
 \ln{\langle R^2 \rangle^{(i)}} \sim A + 2\nu^{(i)}\ln N,
 \end{equation}
 with varying the lower cutoff for the number of steps $N_{{\rm min}}$. The $\chi^2$ values (sums  of squares of normalized deviation from the regression
 line) divided by the number of degrees of freedom DF serves as a test of the goodness of fit (see Tables \ref{tabRSAW}, \ref{tabRRW}, \ref{tabcor}).    
Results of least square fitting give: $\nu_{SAW}=0.593\pm0.001$, $\nu_{RW}=0.499\pm0.001$.
The error bars in Tables  \ref{tabRSAW}, \ref{tabRRW}, \ref{tabcor}  are observed to typically not overlap 
 the known precise values. This is due to the fact that our numerical 
 study involves only relatively short walks, and data for longer walks 
 would be needed to reduce the influence of unfitted corrections to 
 scaling. In addition, the values of the reduced $\chi^2$ are 
 undoubtedly too small, and this is because we do not explicitly take 
 into account the correlations between estimates at different values of 
 $N$ which are introduced by the Rosenbluth-Rosenbluth sampling 
 method.
}

{
\begin{table}[h!]
\begin{center}
\caption {Results of linear fitting of data obtained for $\ln \langle R^2 \rangle^{(4)}_{SAW} $ }
\label{tabRSAW}
  \begin{tabular}{ c  c  c  c }
   \hline
$ N_{{\rm min}} $ & $\nu_{SAW}^{(4)}$ &  $A_{SAW}$ & $\chi^2/DF $  \\ 
\hline
5  & $0.599 \pm 0.002 $ &   $0.094\pm0.002$ & 9.653 \\
10 & $0.596 \pm 0.002$ & $0.105\pm 0.002$ & 4.926\\ 
15 & $0.596 \pm0.001$ & $ 0.113 \pm 0.002$ &  2.912 \\
20 & $0.594 \pm 0.001 $ & $ 0.126\pm0.001$ & 0.993\\
25 & $ 0.593 \pm 0.001$ & $ 0.132\pm0.001$ & 0.822\\
30 & $ 0.593 \pm 0.001 $ &  $0.139\pm 0.001$  &  0.594 \\
\hline
    \end{tabular}
\end{center}
\end{table}

\begin{table}[h!]
\begin{center}
\caption{Results of linear fitting of data obtained for $\ln \langle R^2 \rangle^{(4)}_{RW} $ }
\label{tabRRW}
  \begin{tabular}{ c  c  c  c }
   \hline
$ N_{{\rm min}} $ & $\nu_{RW}^{(4)}$ &  $A_{RW}$ & $\chi^2/DF $  \\ 
\hline
5  & $ 0.495 \pm 0.001 $ &   $  0.189 \pm 0.002$ & 10.259 \\
10 & $  0.497 \pm 0.001$ & $ 0.180 \pm 0.001 $ &  1.021\\ 
15 & $ 0. 499\pm 0.001 $ & $  0.176 \pm 0.001 $ &   0.430\\
20 & $ 0.499 \pm 0.001$ & $ 0.171 \pm 0.001 $ &  0.260 \\
\hline
    \end{tabular}
\end{center}
\end{table}

\begin{table}[h!]
\begin{center}
\caption{Results of linear fitting of data obtained for $\ln \langle \vec{R}_{SAW}^{(4)}\vec{R}_{RW} ^{(4)} \rangle $ }
\label{tabcor}
  \begin{tabular}{ c  c  c  c }
   \hline
$ N_{{\rm min}} $ & $\nu_{cor}^{(4)}$ &  $A_{cor}$ & $\chi^2/DF $  \\ 
\hline
5  & $  0.493 \pm 0.001 $ &   $   -1.144 \pm 0.003$ & 12.459 \\
10 & $   0.495 \pm 0.001$ & $ -1.160 \pm 0.002 $ &  1.940\\ 
15 & $  0.496 \pm 0.001 $ & $  -1.165 \pm 0.001 $ &   0.780\\
20 & $ 0.496 \pm 0.001$ & $ -1.173 \pm 0.001 $ &  0.452 \\
\hline
    \end{tabular}
\end{center}
\end{table}

}

\begin{figure}[b!]
\begin{center}
\includegraphics[width=80mm]{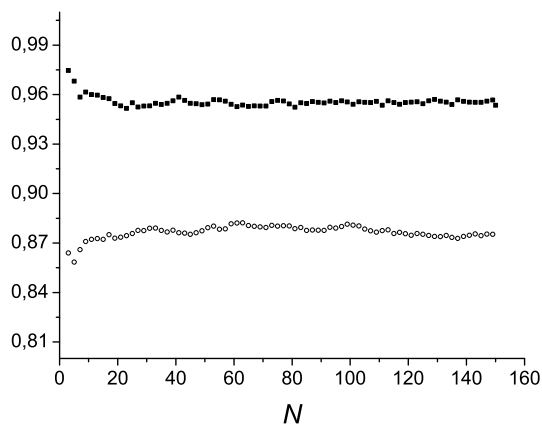}
\end{center}
\caption{The size ratios $g_{SAW}^{(4)}$ (black squares) and $g_{RW}^{(4)}$ (open circles) as functions of $N$.} \label{Ratio}
\end{figure}

When the mutual avoidance is present between the  block chains, they become more extended in space.
To describe this effect quantitatively, we introduce the ratios:
\begin{eqnarray}
&&g_{RW}^{(i)}=\frac{\langle R^2 \rangle_{free}^{RW}}{\langle R^2 \rangle_{RW}^{(i)}},\label{1rr}\\
&&g_{SAW}^{(i)}=\frac{\langle R^2 \rangle_{free}^{SAW}}{\langle R^2 \rangle_{SAW}^{(i)}},\,\,\,\,i=1,\ldots,6.\label{2rr}
\end{eqnarray}
Here, by ``free'' we denote  the averaged end-to-end distances
of the single individual chain of length $N$.
On  Fig. \ref {Ratio},
we give our results for the ratios
$g_{SAW}^{(4)}$ and $g_{RW}^{(4)}$. We can see, that both values are smaller than 1, so that the effective lengths of both block chains are modified and become more extended in space due to mutual avoidance between them. { This effect was described also in previous studies \cite{Tanaka76,Olaj98a}.}

\begin{figure}[t!]
\begin{center}
\includegraphics[width=75mm]{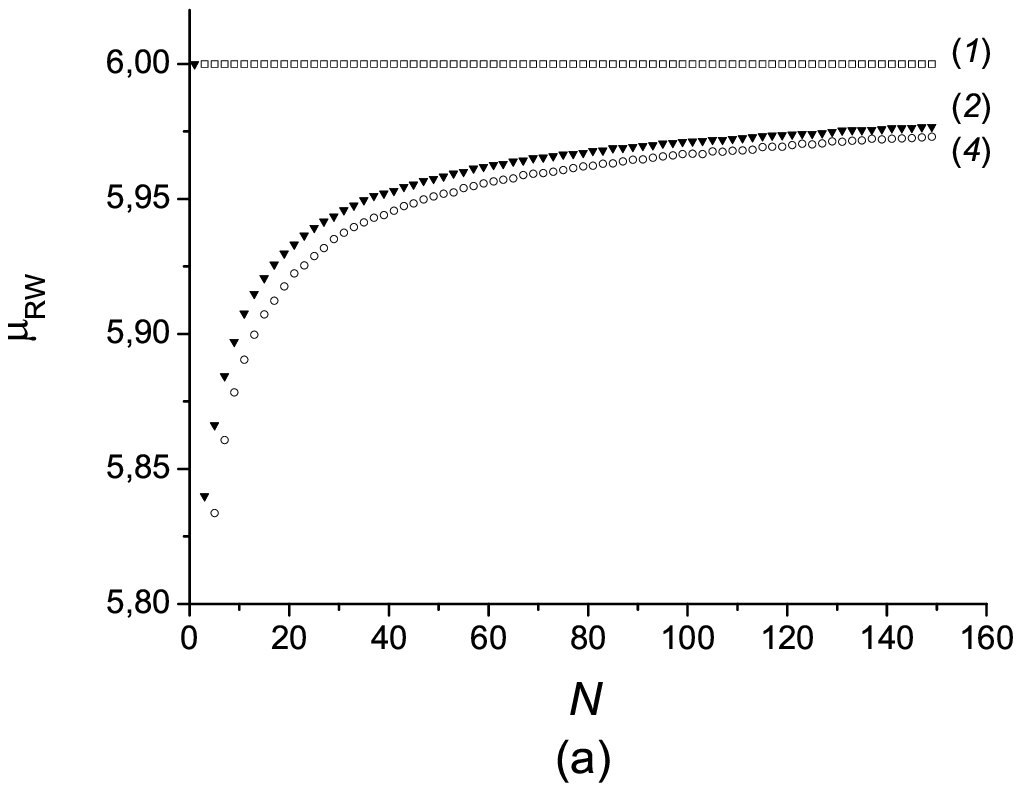}
\includegraphics[width=75mm]{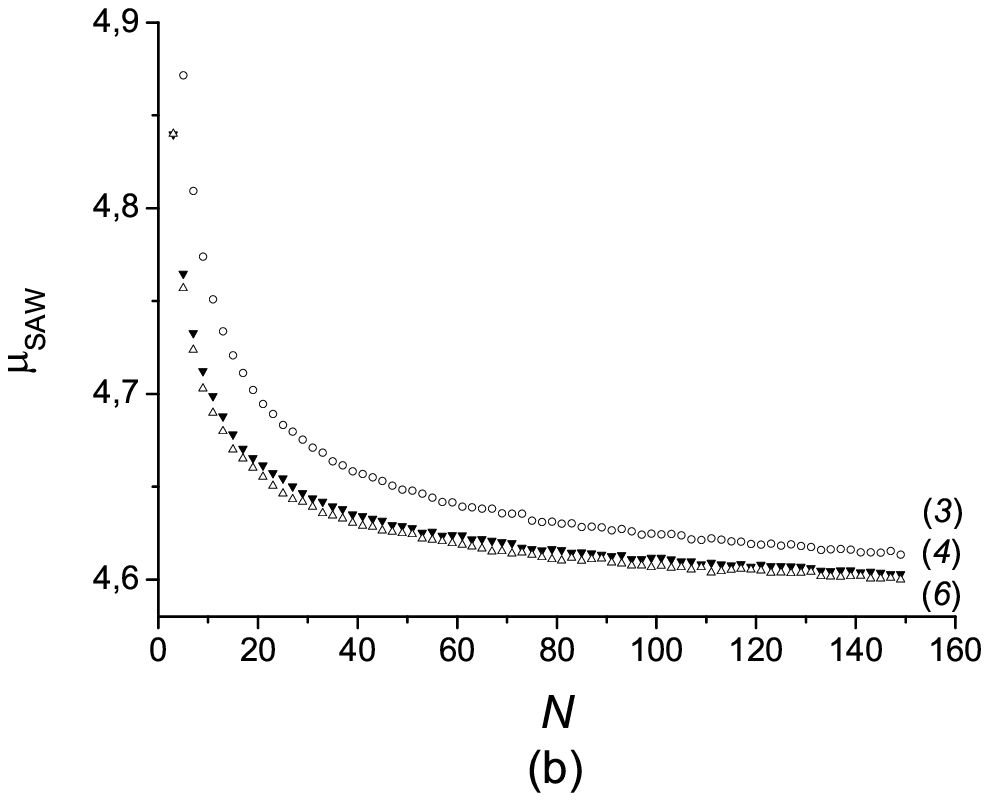}
\end{center}
\caption{The { connectivity constants} of the RW chains (a) and SAW chains (b) considered as blocks of
copolymer chain, as functions of length $N$. } \label{fug}
\end{figure}

{
Such an extension of the effective linear size  of the individual block of copolymer chain caused by interactions with other block
is also confirmed by analyzing the local connectivity constant for each block.}
The data for  averaged { connectivity constants} of both building block chains for the cases (1)-(6) are presented
on Fig. \ref{fug}.  For the case (1),  when we have two RWs without avoidance, we restore $\mu_{RW}^{(1)}=6$; the same holds for individual RW block in the case (3), $\mu_{RW}^{(3)}=6$.
Similarly, for the case (3) and (5), for the SAW block without mutual interaction with other block we approach  the 
 { connectivity  constant} value of single SAW trajectory $\mu_{SAW}^{(3)}=\mu_{SAW}^{(5)}=\mu_{SAW}$,
the most accurate result of which was obtained in the Ref. \cite{Clisby13}: $\mu_{SAW}=4.684039931(27)$.

In all other cases, when mutual interactions between blocks is taken into account, the values of {connectivity constants} are slightly 
modified  as compared with corresponding values of free trajectories. 
This effect becomes less pronounced  with increasing the length of both blocks.
These (even very slight) modifications in the connectivity constants caused by
 presence of mutual avoidance with other bock are actually leading to extension 
of the effective linear size  of individual blocks, analyzed in previous paragraph.



\begin{figure}[b!]
\begin{center}
\includegraphics[width=80mm]{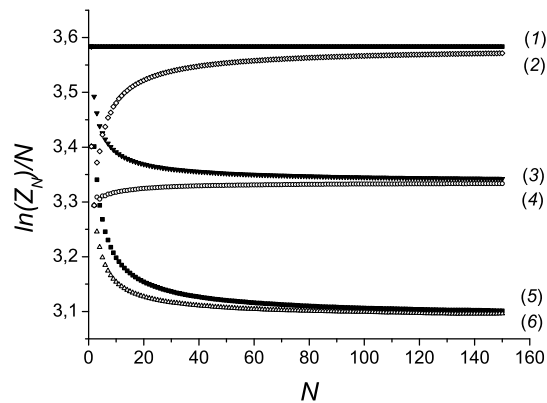}
\end{center}
\caption{Partition sums of the model cases (1)-(6) as functions of the block length $N$.} \label{Statsuma}
\end{figure}

Finally, we consider the partition sum  of diblock copolymer. We rewrite the expression (\ref{stat2})
in the form:
\begin{equation}
\frac{\ln(Z_N)}{N}=a+\ln(\mu_1)+\ln(\mu_2)+(\gamma-1)\ln(N)/N.
\end{equation}
Corresponding results for the cases (1)-(6) are given on Fig. \ref{Statsuma}. As expected, the largest number of configurations
corresponds to the case (1) of two independent random walk trajectories. Taking into account the avoidance of monomers within the block
 and mutual avoidance of different blocks (according to rules (2)-(6)) leads to decreasing of the possible number of configurations.



\section{Discussions}\label{Discus}

The aim of the present paper was to analyze the universal conformational properties of complex macromolecules, consisting of $n$ sequential
blocks of various chemical structure. In general, we consider all the blocks to be of different lengths $L_i$. Of particular interest is the set of
scaling exponents $\gamma(n)$ governing behavior of partition sum (number of possible conformations) as given by Eq. (\ref{stat3}),
and the effective sizes of blocks within the sequence.

Depending on the quality of a solvent, a situation may occur
when some blocks are at the $\theta$-temperature regime, when attractive and repulsive interactions
between monomers compensate each other. These chains { in $d=3$-dimensional case, which is of most interest in 
the real polymer systems}, can be considered as RW trajectories.
The blocks in a regime of good solvent (with repulsive excluded volume interactions between monomers)
  behave as trajectories of SAW, which are not allowed to cross itself.
Also, the interactions
between monomers of different blocks can vanish or be present at some temperatures. Such a situation is closely related to
the case of so-called ternary solutions \cite{schafer1,schafer2}.

As a first step in our analysis, we reconsider the simplest case of diblock copolymer chain (so-called AB-copolymer), consisting of only two blocks.
We studied all possible cases, considering  blocks as RW  or SAW trajectories with and without
 mutual avoiding between them.
In accordance with previous studies \cite{Joanny84,Douglas87,Tanaka76,Olaj98a}, we observed in particular, that the values of scaling exponents governing the effective lengths $\langle R^2_A \rangle$ and $\langle R^2_B \rangle$ of blocks A and B
are not modified by presence of other block. However, when  mutual interaction between blocks is taken into account, the chains become more extended in space.
{ We can propose the following explanation to this fact: the presence of other blocks plays the role of 
spatial hinderness to the given block and can be related to the 
case of a single polymer chain in an environment with structural 
defects. The last problem attracts a considerable 
attention of researchers (see e.g. \cite{Blavatska09} for a review and references therein). 
The main consequence is that the presence of structural defects 
causes the elongation of polymer chain, but does not impact the 
value of exponent $\nu_{SAW}(d)$, unless the concentration of defects 
is above the percolation threshold and the percolation cluster of 
fractal dimension $d_f$ emerges in the system. When one turns 
attention to the empirical Flory formula given below Eq. (1), 
one can intuitively explain this fact. Really, exponent 
$\nu_{SAW}(d)$ is defined only by the space dimension $d$. 
The presence of structural hinderness does not change 
the dimension of space, unless one has percolation cluster 
or any other underlying fractal structure with fractal 
dimension $d_f$. Indeed, simply substituting $d$ by fractal dimension
of percolation cluster gives reliable estimate for 
corresponding $\nu_{SAW}(d_f)$ \cite{Kremer81}.
However, the local crowdedness caused by presence of other blocks 
 does not lead to formation of fractal environment.
Thus, their presence leads to spatial elongation of an individual chain,
as can be seen e.g. by analyzing the size ratios (shown on Figures 5 and 6)
but does not modify the values of size exponents.    
}

This degree of extension  can be quantitatively measured by introducing the size ratio (\ref{1rr}), which allows us to compare the linear size of a block with that of single polymer chain of the same length.
We found the
 analytical estimates for the set of scaling exponents $\gamma$, governing the system of two subsequently connected blocks of different lengths.

Generalizing the direct polymer renormalization group scheme to the system of $n$ subsequently connected blocks
of different lengths $L_i$ in Section \ref{nblock}, we obtained the analytical expressions for scaling exponents $\gamma(n)$ and size ratios $g_i(n)$ of
individual blocks as functions of both the number of blocks and their position along the sequence.
For an illustration, we analyzed in more details the alternating sequence ABABA..., where A and B are correspondingly blocks of the SAW type
and RW type, all of equal length.
Due to interchain contacts with neighboring chains, the effective size measures of blocks are much more extended in space as compared with single polymer chains, and this effect becomes more and more pronounced with increasing the total number of blocks $n$.

\section{Appendix}

Here, we give some details of direct polymer renormalization group scheme, as developed by des Cloiseaux \cite{desCloiseaux}
and generalized by us to the case of diblock copolymers.
The main idea of the method is to eliminate the divergences
of observable quantities, arising in the asymptotic limit of an infinite linear measure,
 by introducing corresponding
renormalization factors, directly connected with physical
quantities. The quantitative values of these observables are
evaluated at the stable fixed point (FP) of the renormalization
group transformation.

The renormalized coupling constants are determined in this process as:
\begin{eqnarray}
&&u_{i,R}(\widetilde{u}_1,\widetilde{u}_2,\widetilde{w}_{1,2}) = - [Z(\widetilde{u}_1,\widetilde{u}_2,\widetilde{w}_{1,2},L_1,L_2)]^{-2} \times\nonumber\\
&&[2\pi L_i \chi_{u_i}(\widetilde{u}_i,\widetilde{w}_{1,2})]^{-2-\epsilon/2} Z_{u_i}(L_1,L_2,L_1,L_2),\\
&&w_{1,2,R}(\widetilde{u}_1,\widetilde{u}_2,\widetilde{w}_{1,2}) = - [Z(\widetilde{u}_1,\widetilde{u}_2,\widetilde{w}_{1,2},L_1,L_2)]^{-2} \times\nonumber\\
&&[2\pi( L_1 \chi_{u_1}(\widetilde{u}_1,\widetilde{w}_{1,2})+L_2 \chi_{u_2}(\widetilde{u}_2,\widetilde{w}_{1,2}))]^{-2-\epsilon/2} \times\nonumber\\
&&Z_{w_{1,2}}(L_1,L_2,L_1,L_2).
\end{eqnarray}
Here, $\chi_{u_i}(\widetilde{u}_i,\widetilde{w}_{1,2})$ are the swelling factors for the corresponding end-to-end distances:
 \begin{equation}
 \langle R^2_i\rangle = dL_i \chi_{u_i}, i=1,2,
 \end{equation}
  and $Z_{u_1}$, $Z_{u_2}$, $Z_{w_{1,2}}$ are so-called two-polymer partition functions.

\begin{figure}[h!]
\begin{center}
\includegraphics[width=60mm]{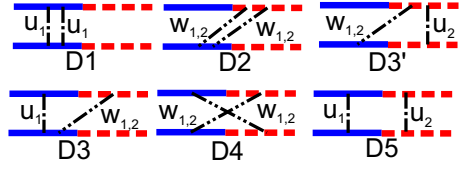}
\end{center}
\caption{Schematic representation of  contributions into the two-polymer function $ \widetilde{Z}(L_1,L_2,L_1,L_2)$. Here,
  solid and  dash lines represent two blocks of copolymer chains, and the dot-dashed lines represent the interaction between chains,
   governed by corresponding coupling constants.} \label{fig:2}
\end{figure}

We need to calculate additionally a two-polymer partition function, that can be presented as:
\begin{eqnarray}
&&Z(L_1,L_2,L_1,L_2) = -u_1Z^2-u_2Z^2-w_{1,2}Z^2+ \widetilde{Z}(L_1,L_2,L_1,L_2)
\end{eqnarray}
where $ \widetilde{Z}(L_1,L_2,L_1,L_2)$ is given by a sum of all contributions, presented diagrammatically on  Fig. \ref{fig:2}. The analytical expressions,
corresponding to presented diagrams, read:
\begin{eqnarray}
&&D1=u_iL^2_i\widetilde{u}_i\left(\frac{2}{\epsilon}-2\ln(2)+1/2\right),\\
&&D2=w_{1,2}L_1L_2\widetilde{w}_{1,2}\left(\frac{2}{\epsilon}+\frac{1}{2}-\frac{L_1}{2L_2}\,\ln  \left( \frac {L_1+ L_2}{L_1}\right)\right.\nonumber\\
&&\left.-\frac{L_2}{2L_1}\,\ln  \left( \frac {L_1+ L_2}{L_1} \right)
-\ln  \left( \frac { \left(L_1+ L_2\right) ^{2}}{ L_1L_2}\right)\right),\\
&&D3(D3')=\frac{u_iw_{1,2}}{(2\pi)^{d/2}}\left(-\frac{\left( 2\,{ L_1}+{L_2} \right) ^{2}}{2}\,\ln  \left( 2\, L_1+L_2 \right)+\right.\nonumber\\
&&+\ln  \left( {L_1}+{L_2} \right)  \left( {L_1}+{L_2} \right)  \left( 2\,{L_1}+{L_2} \right) -\frac{{{L_2}}^{2}}{2}\,\ln  \left( { L_2} \right)\nonumber\\
 &&\left.-{L_1}\,{L_2}\,\ln  \left({ L_2} \right) +2\,{{L_1}}^{2}\ln  \left( 2 \right)\right),\\
&&D4= \frac{w^2_{1,2}}{(2\pi)^{d/2}}\left(\left( { L_1}+2\,{L_2} \right) ^{2}\ln  \left( {L_1}+2\,{L_2} \right) \right.\nonumber\\
&&-{{L_1}}^{2}\ln  \left( {L_1} \right)+\left( 2\,{L_1}+{L_2} \right) ^{2}\ln  \left( 2\,{L_1}+{L_2} \right)\nonumber\\
 &&-4\, \left( {L_1}+{L_2} \right) ^{2}\ln  \left( {L_1}+{L_2} \right) \nonumber\\
&&\left.-4\,\ln  \left( 2 \right) \left( {{L_1}}^{2}+{L_1}\,{L2}+{{L_2}}^{2} \right) -{{L_2}}^{2}\ln  \left( {L_2} \right)\right),\\
&&D5= \frac{u_1u_2}{(2\pi)^{d/2}}\left( (2L_1+L_2)^2\ln(2L_1+L_2)\right.\nonumber\\
&&+(L_1+2L_2)^2\ln(L_1+2L_2)-(L_1+L_2)^2\ln(L_1+L_2)\nonumber\\
&&\left.-2(L_1+L_2)^2\ln(2)-L_1^2\ln\left(\frac{L_1}{4}\right)-L_2^2\ln\left(\frac{L_2}{4}\right)\right).
\end{eqnarray}
The diagrams $D1$ and $D3$ have a pre-factor $2$, diagram $D2$ has a pre-factor $4$ and the $D4$ and $D5$ are accounted only once.
Note also, that we need to take into account only the terms containing the poles $\epsilon^{-1}$.

The flows of the renormalized coupling constants are
governed by $\beta$ functions defined according to
\begin{eqnarray}
&&\beta_{u_{i,R}} = 2L_i\frac{\partial u_{i,R}}{\partial L_i},\\
&&\beta_{w_{1,2,R}} = 2\sum_{i=1}^2L_i\frac{\partial w_{1,2,R}}{\partial L_i}\label{beta}.
\end{eqnarray}
We receive:
\begin{eqnarray}
&&\beta_{u_{i,R}} =\epsilon u_{i,R}-8u^2_{i,R}=0, \label{b1}\\
&&\beta_{w_{1,2,R}} =\epsilon w_{1,2,R}-\frac{(L_1+L_2)^2}{L_1L_2}w^2_{1,2,R} -2w_{1,2,R}(u_{1,R}+u_{2,R}). \label{b2}
\end{eqnarray}
The FPs of the renormalization group transformations, allowing us to obtain the quantitative values of
scaling exponents (\ref{gamma}), (\ref{nu}), are
defined as common zeros of functions (\ref{b1}), (\ref{b2}).


\section*{References} 

\begin{thebibliography}{50}

\bibitem{Hadjichristidis03}
 Hadjichristidis N, Pispas S and  Floudas G 2003 \textit{Block Copolymers: Synthetic Strategies, Physical Properties, and Applications}
(Inc: John Wiley \& Sons).

\bibitem{Matsen96}
Matsen M W and  Bates F S 1996 {\it Macromolecules}  {\bf 29}
7641 

\bibitem{Bates99}
 Bates F S and  Fredrickson G H 1999 {\it Phys. Today} {\bf 52} 32

\bibitem{Mai12}
 Mai Y and  Eisenberg A 2012 {\it Chem. Soc. Rev.}  {\bf 41}  5969 

\bibitem{Jackson10}
Jackson E A  and  Hillmyer M A 2010 {\it ACS Nano} {\bf 4}   3548

\bibitem{Dami17}
Dami S,  Abetz C,  Fischer B, Radjabian M,
Georgopanos P and Abetz V 2017 {\it Polymer} {\bf 126}  376 









\bibitem{OssRonen12}
Oss-Ronnen L,  Schmidt J, Abetz V,  Radulescu A, Cohen Y,
and Talmon Y 2012 {\it Macromolecules} {\bf 45}  9631 

\bibitem{Meng09}
Meng F,  Zhang Z, and  Feijen J 2009 {\it Biomacromolecules} {\bf 10}  197 

\bibitem{Ruiz06}
Ruiz R et al. 2008 {\it Science} {\bf 321} 936 

\bibitem{Bates01}
Bates F S,  Fredrickson G H,  Hucul D, and Hahn S F 2001
{\it AIChE J.} {\bf 47} 762

\bibitem{Schutz05}
 Schultz A J,  Hall C K, and  Genzer J 2005 {\it Macromolecules} {\bf 38} 3007 

\bibitem{Uhrig02}
Uhrig D and  Mays J W 2002   {\it Macromolecules}  {\bf 35} 7182 

\bibitem{Feng17}
Feng H, Lu X,  Wang W,  Kang N G, and  Mays J W 2017 {\it Polymers} {\bf 9} 494 

\bibitem{Bates12}
Bates F S et al. 2012 {\it Science} {\bf 336}  434 




\bibitem{Joanny84}
 Joanny J,  Leibler L., and Ball R 1984 {\it  J. Chem. Phys.} {\bf 81} 4640
 
 \bibitem{Douglas87}
  Douglas J F  and  Fried K F (1987) {\it J. Chem. Phys.} {\bf 86} 4280
 
 \bibitem{McMullen89}
  McMullen W E,  Freed K F  and  Cherayil B J 1989 {\it Macromolecules} {\bf 22} 1853
  
  
  \bibitem{Tanaka76}
   Tanaka T,  Kotaka T and  Inagaki H 1976 {\it Macromolecules} {\bf 9} 561
   
   \bibitem{Tanaka77}
    Tanaka T,  Kotaka T,  Ban K,  Hattori M and  Inagaki H 1977 {\it Macromolecules} {\bf 10} 960
    
    \bibitem{Tanaka79}
    Tanaka T,  Omoto M and Inagaki H 1979  {\it Macromolecules} {\bf  12} 146
    
    \bibitem{Molina94}
     Molina L A,  Rodriguez A L, and  Freire J J 1994 {\it Macromolecules} {\bf 27} 1160
     
     \bibitem{Olaj98a}
      Olaj O F,  Neubauer B and  Ziferer G 1998 {\it Macromol. Theory Simul.} {\bf 7} 171
      
      
       \bibitem{Sdranis91}
        Sdranis Y S and  Kosmas M K 1991 {\it Macromolecules} {\bf 24} 1341
        
          
       \bibitem{Olaj98b}
      Olaj O F,  Neubauer B and  Ziferer G 1998 {\it Macromol. Theory Simul.} {\bf 7} 181
      
      \bibitem{Olaj98c}
       Neubauer B,  Ziferer G and Olaj O F  1998 {\it Macromol. Theory Simul.} {\bf 7} 189
       
       
       \bibitem{Theodorakis12}
       Theodorakis  P E and  Fytas N G  2012 {\it J. Chem. Phys. } {\bf 136} 094902 



 





\bibitem{deGennes} de Gennes P G 1979  \textit{Scaling Concepts in Polymer Physics } (Ithaca, NY: Cornell 
University Press). 

\bibitem {desCloiseaux}  des Cloizeaux J and Jannink G 1990 \textit{Polymers in Solutions: Their 
Modelling and Structure} (Oxford: Clarendon Press).

\bibitem{Duplantier87}
 Duplantier B and Saleur H 1987 {\it Phys. Rev. Lett.} {\bf 59} 539

\bibitem{Duplantier86}
Duplantier B 1986 {\it Europhys. Lett.} {\bf 1} 491

\bibitem{Duplantier86a}
 Duplantier B 1987 {\it  J. Chem. Phys.} {\bf 86} 4233



\bibitem{Properties}
 Mark J E (ed.) 1996
\textit{Physical Properties of Polymers Handbook} {NY: AIP Press}

\bibitem{Polymerization}
 Dotson N A,  Galvan R,  Laurence R L, and  Tirrell M 1996 \textit{Polymerization process modeling} {	NY: John Wiley and Sons}

 


\bibitem{Nienhuis82}
 Nienhuis B 1982 {\it Phys. Rev. Lett.} {\bf 49} 1062 

\bibitem{Clisby16} 
  Clisby N and D\"nweg B 2016 {\it Phys. Rev. E} {\bf 94} 052102   




\bibitem{Nienhuis84}
Nienhuis B 1984  {\it J. Stat. Phys.} {\bf 34} 731 

\bibitem{Clisby17}
Clisby N 2017 {\it J. Phys. A: Math. Theor} {\bf  50} 264003








\bibitem{schafer1}
Sch\"afer L and  Kapeller C 1985 {\it J. Phys. (Paris)} {\bf 46} 1853;
1990 {Colloid Polym. Sci.} {\bf 268} 995

\bibitem{schafer2}
 Sch\"afer L,  Lehr U, and
 Kapeller C 1991 {\it J. Phys. I} {\bf 1} 211 



\bibitem{Holovatch97}
von Ferber C and  Holovatch Yu 1997 {\it Phys. Rev. E} {\bf 56 }  6370



\bibitem{Duplantier88}
Duplantier B 1988 {\it Commun. Math. Phys.} {\bf 117} 279 

\bibitem{Duplantier88a}
Duplantier B and  Kwon K - H 1988 {\it Phys. Rev. Lett.} {\bf 61} 2514

\bibitem{Lawler82}
Lawler G H 1980 {\it Duke Math J.} {\bf  47} 655

\bibitem{Lawler90}
Burdzy K, Lawler G F and Polaski T 1989 {\it J. Stat. Phys.} {\bf 56} 1 

\bibitem{Li90}
Li B and  Sokal A D 1990 {\it J. Stat. Phys.} {\bf 61} 723






\bibitem{Rosenbluth55}  Rosenbluth M N  and  Rosenbluth A W 1955  {\it J. Chem. Phys.} {\bf 23}   356

\bibitem{Clisby13} Clisby N 2013 {\it J. Phys. A: Math. Theor.} {\bf 46} 245001 


\bibitem{Blavatska09}
 Blavatska V and  Janke W 2009 {\it J. Phys. A} {\bf 42} 015001 

\bibitem{Kremer81}
Kremer K 1981 {\it Z. Phys. B} {\bf 45} 149

\end{thebibliography}
\end{document}